\documentstyle[12pt,epsfig]{article}
\def\be{\begin{equation}}
\def\ee{\end{equation}}

\def\lsim{\raise0.3ex\hbox{$<$\kern-0.75em\raise-1.1ex\hbox{$\sim$}}}
\def\gsim{\raise0.3ex\hbox{$>$\kern-0.75em\raise-1.1ex\hbox{$\sim$}}}

\begin{document}
\thispagestyle{empty}

%\noindent April 27, 2001 \hfill draft

%\vskip 1.5 cm

\centerline{\large{\bf Cluster Percolation and}}

\medskip

\centerline{\large{\bf Pseudocritical Behaviour in Spin Models}}

\vskip 1.0cm

\centerline{\bf Santo Fortunato and Helmut Satz}

\bigskip

\centerline{Fakult\"at f\"ur Physik, Universit\"at Bielefeld}
\par
\centerline{D-33501 Bielefeld, Germany}

\vskip 1.0cm

\noindent

\centerline{\bf Abstract:}

\medskip

The critical behaviour of many spin models can be equivalently
formulated as percolation of specific site-bond clusters. In the
presence of an external magnetic field, such clusters remain
well-defined and lead to a percolation transition, even though the
system no longer shows thermal critical behaviour. We investigate the
2-dimensional Ising model and the 3-dimensional $O(2)$ model by means
of Monte Carlo simulations. We find for small fields that the line of
percolation critical points has the same functional form as the line of
thermal pseudocritical points.

\vskip 1.5cm

Percolation theory \cite{stauff,grimm} provides in many cases an
elegant interpretation of the mechanism of second order phase
transitions: the existence of an infinite spanning cluster represents
the new order of the microscopic constituents of the system due to
spontaneous symmetry breaking.

In the Ising model, a rigorous correspondence between thermal critical
behaviour and percolation was established by Coniglio and Klein
\cite{conkl}; the clusters are constructed by linking like-sign spins
through a temperature-dependent bond probability $p=1-\exp(-2J/kT)$,
where $J$ is the Ising coupling and $T$ the temperature. This result
has been recently extended to a wide class of models, from continuous
spin Ising-like models \cite{bia,sasa} to $O(n)$ models \cite{tere}.
In all cases the system was considered in the absence of any external
field. The reason for this is clear: by introducing an external field
$H$, we explicitly break the symmetry of the Hamiltonian of the
system, thus eliminating the thermal critical behaviour of the model.
None of the thermodynamic potentials exhibits discontinuities of any
kind, since the partition function is analytical for $H \neq 0$.

On the other hand, the Coniglio-Klein clusters can be built as well
when $H{\neq}0$. Because of the field, the system has a non-vanishing
magnetization $m$ parallel to the direction of $H$ for any finite value
of the temperature $T$. For $T\rightarrow\infty$, $m{\rightarrow}\,0$,
and for $T=0$, $m=1$. This suggests that for a fixed value of $H$, the
clusters will start to form an infinite network at some temperature
$T_p(H)$. Varying the field $H$, one thus obtains a curve $T_p(H)$ in
the $T-H$ plane, the Kert\'esz line \cite{kertesz}.

The Kert\'esz line specifies the usual percolation threshold, just as
in the case $H=0$, and the percolation variables exibit the usual
singularities, leading to a set of critical exponents. In particular,
the percolation strength, which is the relative size of the percolation
cluster compared to the total volume of the system, remains the order
parameter of the percolation transition.

The existence of such a line of genuine critical behaviour also for
$H{\neq}0$ could provide a criterium to define different phases and the
transition between them in a more general sense \cite{satz}. For this
purpose, it is important to find out whether there is any relationship
between the geometrical singularities at the Kert\'esz line and thermal
properties of the system along this line; apparently, rather little
is known about such connections \cite{adler}-\cite{campi2}. 

In the present work we want to address this aspect in more detail. In
particular, we shall investigate the relation between the Kert\'esz
line and the so-called "pseudocritical" or "crossover" line, defined
by the values of the temperature at which the magnetic susceptibility
peaks as function of the field. In general, the two lines cannot coincide,
since when $H \to \infty$, the Kert\'esz line leads to a finite $T$, 
while for the pseudocritical line this results in $T \to \infty$; we
return to this point later on. We shall here study two spin models, the
Ising model in two dimensions and the $O(2)$ model in three dimensions,
and calculate in both cases the functional form of the Kert\'esz line
and the pseudocritical line in the limit of a small external field.

The Hamiltonian of the models at study can be written in the following
general form:
\be
\label{ham}
{\beta}{\cal H}\,=\,-\frac{J}{kT}\sum_{\langle{i,j}\rangle}{\bf s_i
  s_j}-\frac{H}{kT}\sum_i{\bf s_H s_i},
\ee
where $\bf s_i, s_j$ are two-dimensional unit vectors for $O(2)$ and
simple scalars (${\pm}1$) for the Ising model; similarly, $\bf s_H$ is a
unit vector (or a scalar) in the direction of the external field. We now first
consider the percolation problem.

\begin{figure}[h]
\begin{center}
\epsfig{file=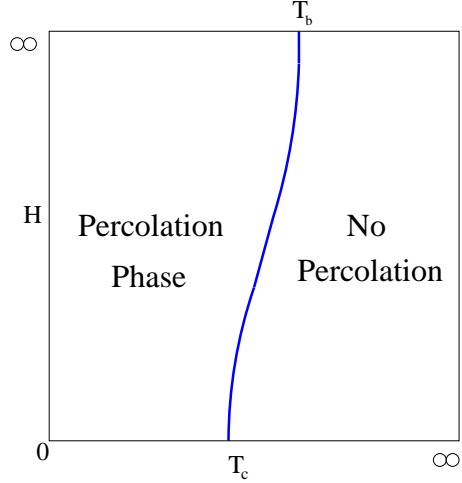,  width=6cm}
\caption{Schematic behaviour of the Kert\'esz line for the Ising model}
\label{kert}
\end{center}
\end{figure}

In Fig.\ \ref{kert} we show schematically the Kert\'esz line for the
Ising model. For a vanishing external field, $H=0$, the Coniglio-Klein
cluster definition with
\be
p_{i,j} \equiv p_b = 1 - \exp\{-2J/kT\}
\label{CK}
\ee
as bond weight between two adjacent spins $i,j$ assures that we recover
the usual thermal threshold of the Ising model. For $H \rightarrow
\infty$, all lattice spins will be aligned with the field at any
temperature $T$. However, the bonds between adjacent spins will be
occupied according to the bond weight $p_b$, so that now the site-bond
problem turns into pure bond percolation. The percolation transition
will then take place for that value of the temperature $T_b$ for which
the probability $p_{ij}$ equals the critical density $p_b(d)$ of random
bond percolation in $d$ dimensions,
\be
\label{limva}
p_b(d) = 1-\exp\{-2J/kT_b\},
\ee
i.e., for
\be
T_b = {-2J \over k\,\log[1-p_b(d)]}
\label{temp}
\ee
These arguments can simply be repeated for the Kert\'esz lines
corresponding to $O(n)$ spin models.

As mentioned above, each point of the Kert\'esz line is a standard
percolation point. At $H=0$, the critical exponents of the percolation
variables coincide with the thermal critical exponents of the
magnetization transition (Ising, $O(n)$). However, for any $H{\neq}0$,
they are predicted to switch into the (different) universality class of
random percolation in the same dimension. We will verify this for each
of the percolation points we shall determine.

\vskip0.1cm
\noindent

Turning now to the pseudocritical thermal behaviour of the models
considered, we define the reduced temperature $t$ and external field
$h$ measured relative to the spin-spin coupling strength
\be
\label{reduce}
t=\frac{T-T_c}{T_c}, \hskip 1cm h=\frac{H}{J}.
\ee
For spin systems, the renormalization group approach leads to the
following expression of the magnetization $m$ when $t$,~$h\,{\ll}\,1$
\be
\label{miak}
m\,=\,h^{1/\delta}f_G(t/h^{1/\beta\delta}),
\ee
where $f_G$ is a scaling function and $\beta$, $\delta$ are the critical
exponents for the magnetization transition at $h=0$. We can rewrite
Eq.\ (\ref{miak}) in the form
\be
\label{rewr}
m\, =
\,t^{\beta}({t/h^{1/\beta\delta}})^{-\beta}f_G(t/h^{1/\beta\delta})
\,=\,
t^{\beta}F_G(t/h^{1/\beta\delta})
\ee
with a new scaling function $F_G$; the entire dependence on the field
$h$ is now put into the function $F_G$.

The susceptibility $\chi$ is given by
\be
\label{chi}
\chi\, \equiv \,
\left(
\frac{\partial{m}}{\partial{H}} \right)_T
\,\propto\,t^{\beta}F_G^{\prime}(x) th^{-1/\beta\delta-1},
\ee
where $F_G^{\prime}(x)$ is the derivative of the function $F_G$ with
respect to its argument $x=t/h^{1/\beta\delta}$. This time we move the
dependence on the reduced temperature $t$ into a new function $\cal F$
of the argument $x$ and obtain
\be
\label{newchi}
\chi\,\propto\,h^{1/\delta-1}{\cal F}(t/h^{1/\beta\delta})
\ee
To find the equation of the pseudocritical line, we have to determine
the temperature $t_{\chi}(h)$ at which the susceptibility $\chi$ peaks
for a given value $h$ of the field. Hence the derivative of $\chi$ with
respect to the reduced temperature $t$ must vanish at $t=t_{\chi}$,
\be
\label{derchi}
\frac{\partial{\chi}}{\partial{t}}\Big\vert_{t=t_{\chi}}\,=\,0\,\,\rightarrow\,\,
h^{1/\delta-1}
{\cal F}^{\prime}(x)|_{t=t_{\chi}}(1/h^{1/\beta\delta})\,=\,
h^{1/\delta-1-1/\beta\delta}{\cal F}^{\prime}(x)|_{t=t_{\chi}}\,=\,0;
\ee
here ${\cal F}^{\prime}(x)|_{t=t_{\chi}}$ is the derivative of the
function ${\cal F}$ with respect to its argument $x$ calculated
at $t=t_{\chi}$. The derivative
${\partial{\chi}}/{\partial{t}}|_{t=t_{\chi}}$ can be zero only
if ${\cal F}^{\prime}(x)|_{t=t_{\chi}}=0$. That will occur at some value
$x_{\chi}$ of the argument $x=t_{\chi}/h^{1/\beta\delta}$, which
provides the relation between $t_{\chi}$ and $h$
\be
\label{pseudo}
t_{\chi}/h^{1/\beta\delta}=x_{\chi}
\,\,\rightarrow\,\,t_{\chi}\,\propto\,h^{1/\beta\delta}.
\ee
The procedure is obviously independent of the value of the field $h$,
so that as long as $h$ is small, the pseudocritical line is described
by a simple power law.

We stress that, for $h\rightarrow\infty$, the temperature of the 
susceptibility peak diverges, $t_{\chi}\rightarrow\infty$, whereas we 
have seen that the Kert\'esz line has a finite endpoint, given by 
Eq.\ (\ref{limva}). Hence the two curves, which start from the same point
but tend to two different limits, must certainly differ for sufficiently 
large values of $h$. We want to investigate here, however, 
what happens in the vicinity of the thermal critical
point of the model, i.e. for small fields ($h\,{\approx}\,0$).

In order to study the percolation transition we need to redefine each
given spin configuration as a cluster configuration, by grouping all
spins of the lattice in clusters. We do this with the help of the
algorithm of Hoshen and Kopelman \cite{kopelman}, using free boundary
conditions for the cluster identification and assuming that a cluster
percolates if it connects each pair of opposite sides (faces in 3d) of
the lattice.

After measuring the size of all clusters, i.e., the number of sites
in each cluster, we evaluate the following two quantities:
\begin{itemize}
\item{The {\it average cluster size } $S$,
\be
S\equiv \frac{\sum_{s} {{n_{s}s^2}}}{\sum_{s}{n_{s}s}}~,
\ee
where $n_{s}$ is the number of clusters of size $s$; the sums exclude
the percolating cluster.}
\item{The {\it percolation strength } $P$,
\be
P=\frac{\mbox{\rm size of the percolating cluster}}
            {\mbox{\rm total number of lattice sites}},~
\ee
which is the order parameter of the percolation transition.}
\end{itemize}
\noindent
In the infinite volume limit, the percolation variables are described
by simple power laws
\be
\label{pow}
P\,\propto\,(T_p-T)^{\beta_p}~~~(T<T_p), \hskip 1cm
S\,\propto\,|T-T_p|^{-\gamma_p}
\ee
sufficiently close to the critical threshold $T_p$.

In order to determine the critical point of the percolation transition,
we exploit the properties of a further very useful variable. For
finite lattice size, there can be spanning clusters even for
temperatures $T>T_p$, and there are spin configurations at temperatures
$T<T_p$ below the critical threshold without such clusters.
The probability of finding a spanning cluster on a finite lattice of
linear dimension $L$ at a temperature $T$ is a well-defined function,
the {\it percolation cumulant} $\Pi(T,L)$ \cite{bind}. For $T \simeq
T_p$ and large $L$, it behaves as
\be
\label{Pprob}
\Pi(T,L)\,=\,\Phi\Big[\Big(\frac{T-T_p}{T_p}\Big)\,L^{1/{\nu_p}}\Big].
\ee
where $\nu_p$ is the critical exponent governing the divergence of the
percolation correlation length. The function $\Pi(T,L)$ is not a genuine
percolation variable, since it has a non-trivial meaning only on finite
lattices. On an infinite lattice it reduces to a step function: it is
zero for $T>T_p$ and unity for $T<T_p$. Nevertheless, its particular
features make it a powerful tool to extract information about the
critical properties of percolation. In particular, at $T=T_p$,
$\Pi(T_p,L)=\Phi(0)$ for any value of $L$. That means that if we
calculate the percolation cumulant as a function of $T$ for different
lattice sizes, all curves $\Pi(T,L)$ will cross at the critical
temperature $T_p$. Moreover, if we plot the different $\Pi(T,L)$ as a
function of the variable $X=[(T-T_p)/T_p] L^{1/{\nu_p}}$, the
resulting functions $\Phi(X)$ must coincide for all lattice sizes. To
define the scaling function $\Phi(X)$, one also has to know the value
of the exponent $\nu_p$. We can therefore determine $\nu_p$ by looking
for the best scaling form of the percolation cumulant curves.

Let us now first consider the results obtained for the 2D Ising model.
Our aim is to determine the functional form $t_p(h)$  describing the
Kerte\'sz line. In order to compare it to the pseudocritical line, we
have to perform simulations at small values of the reduced field $h$.
We consider the following four values: $h$= 0.002, 0.001, 0.0005 and
0.00025. The Monte Carlo algorithm used in the simulations is the Wolff
cluster update extended to the case of a non-vanishing external field
\cite{nieder}. For each iteration we have measured the energy
$\epsilon$ of the configuration, its magnetization $m$, and the
percolation variables $S$ and $P$. To determine the percolation
critical point with a good accuracy, we work with rather large
lattices: $300^2$, $500^2$, $700^2$, $1000^2$. This is helpful also
for another reason. Since the field is quite weak, the scaling
behaviour of both thermal and percolation variables could be perturbed
by the ``near-by" case of the system without field, which could affect
the values of the critical exponents. Such perturbation becomes smaller
the larger the size of the lattice.
\vspace{-0.5cm}
\begin{figure}[htb]
\begin{center}
\epsfig{file=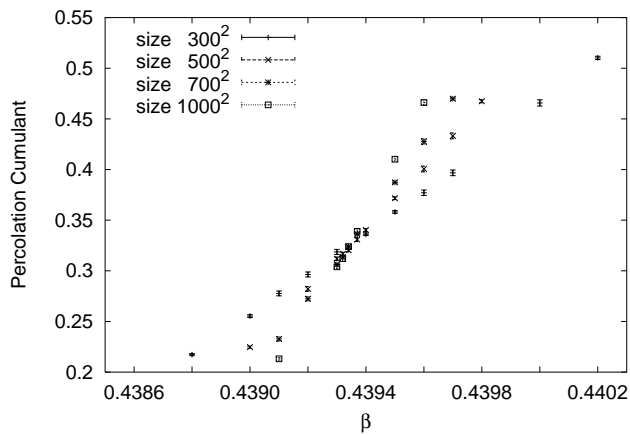,width=9.7cm}
\caption{\label{cum}{Percolation cumulants for four lattice sizes;
$\beta=J/kT$, $h$=0.00025.}}
\end{center}
\end{figure}

\begin{figure}[htb]
\begin{center}
\epsfig{file=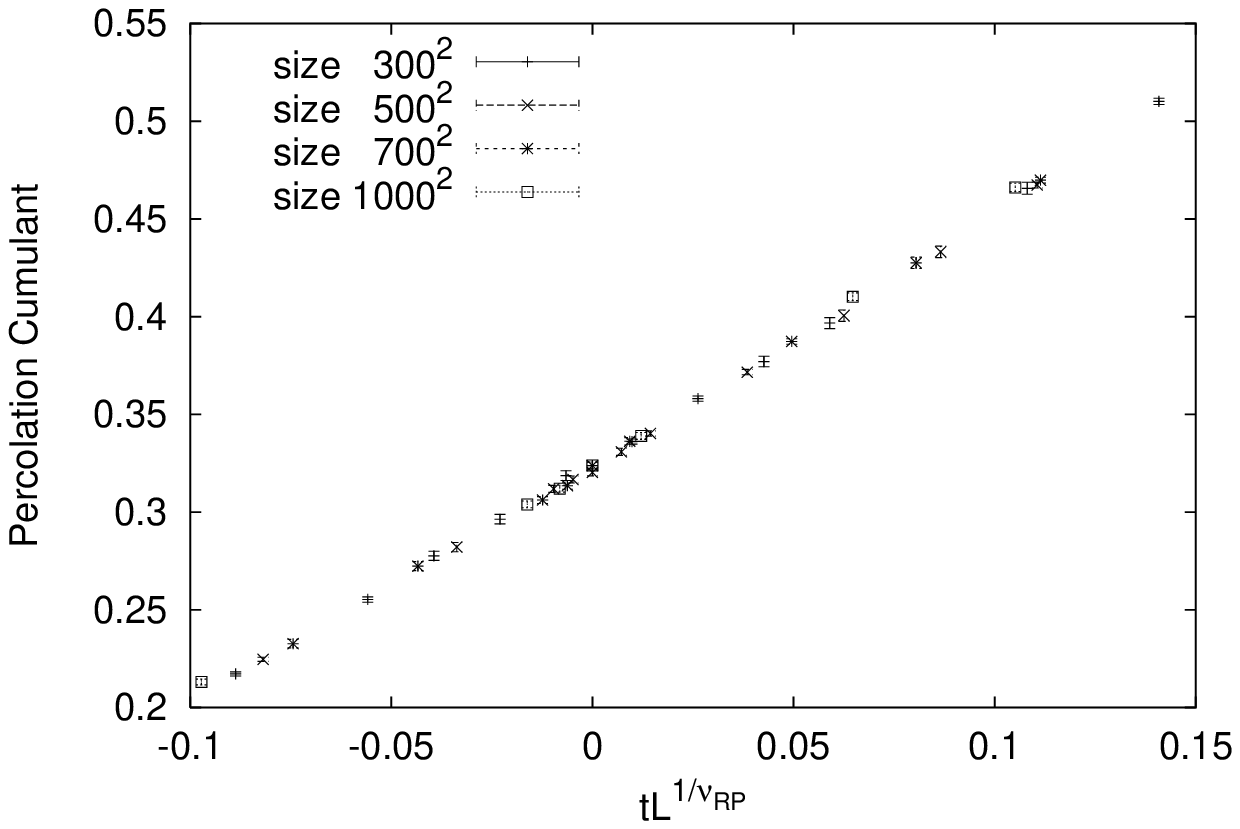,width=11.5cm}
\caption{\label{scalcump}{The rescaled percolation cumulants of
Fig.\ \ref{cum} with $J/kT_p=0.43933$ and the 2-dimensional random
percolation exponent $\nu_{RP}=4/3$.}}
\vskip0.5cm
\epsfig{file=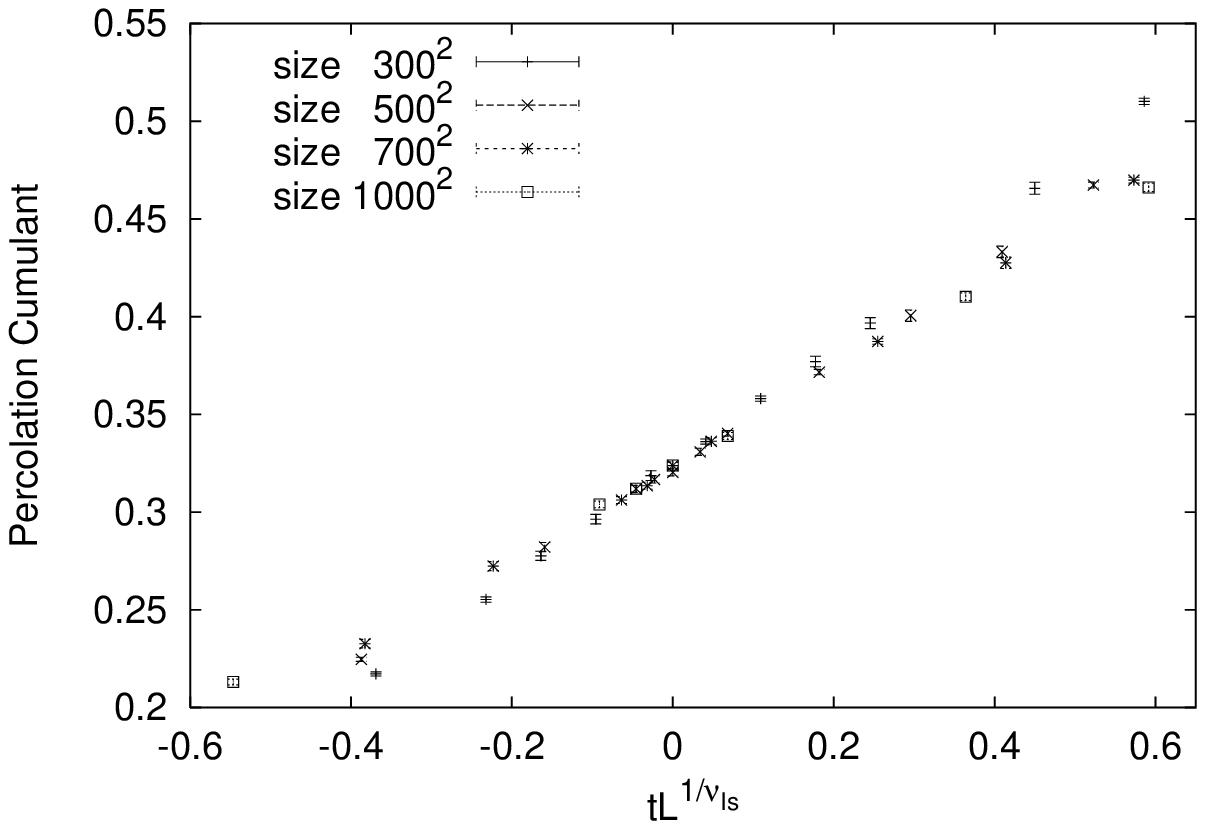,width=11.5cm}
\caption{\label{scalcumis}{The rescaled percolation cumulants
of Fig.\ \ref{cum} with $J/kT_p=0.43933$ and the
2-dimensional Ising exponent $\nu_{IM}=1$.}}
\end{center}
\end{figure}

Fig.\ \ref{cum} shows the percolation cumulants as a funtion of
$\beta=J/kT$ for different lattice sizes, at the smallest field we
have studied, $h=0.00025$. The curves cross clearly at the same point
($\beta_p=0.43933$).
To check the universality class of the percolation exponents
in this case, we rescale the percolation cumulants as discussed
in Section 3. Using $J/kT_p=0.43933$, we consider two options for
the critical exponent $\nu_p$, the 2-dimensional random percolation
value $\nu_{RP}=4/3$ and the 2D Ising exponent $\nu_{IM}=1$. The results
are shown in Figs.\ \ref{scalcump} and \ref{scalcumis}; they clearly
indicate that the critical behaviour at the Kert\'esz line is determined
by the random percolation exponent $\nu_{RP}$, as expected.

\vspace{0.2cm}
\clearpage

\begin{figure}[htb]
\begin{center}
\epsfig{file=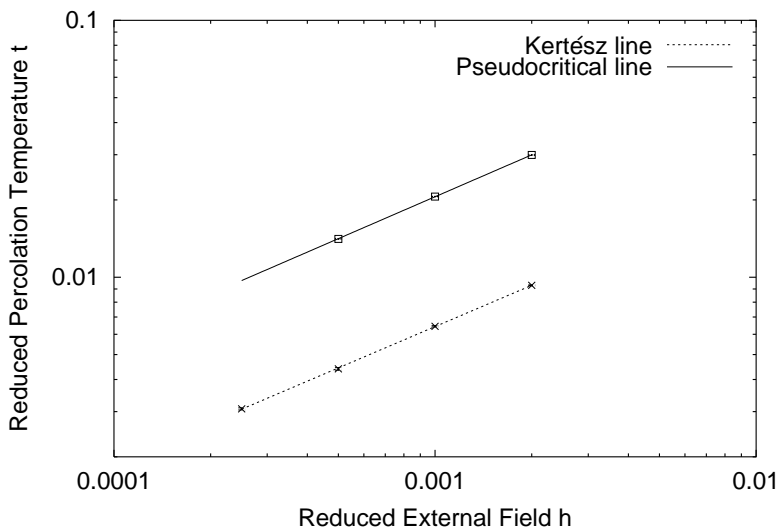,width=12cm}
\caption{\label{ker}{Kert\'esz line and pseudocritical line for the 2D
Ising model.}}
\end{center}
\end{figure}

Using a standard finite-size scaling analysis of the average cluster
size $S$ near the critical point, we obtain for the ratio of the
exponents $\gamma_p/\nu_p=1.80(2)$, in accord with the value predicted
from the random percolation exponents (129/72=1.7916) and not with that
from the Ising model (7/4=1.75).
Repeating the percolation cumulant study for the other three values of
$h$, we find the critical percolation temperatures listed in Table
\ref{thre} and plotted in Fig.\ \ref{ker}. In the Figure, the
reduced percolation temperature $t_p = (T_p-T_c)/T_c$ is shown as
function of the external field; here $T_c$ is the critical temperature
of the Ising model without field
($J/kT_c=(1/2)\log(\sqrt{2}+1)=0.44068679$).
On the logarithmic scale of the plot the data points fall remarkably
well on a straight line. We thus conclude that, for the Kert\'esz line
of the 2D Ising model,

\be
\label{miake}
t_p = a_p h^{\kappa}, \hskip 0.5cm {\rm for} ~ h\,{\rightarrow}\,0,
\ee
where $a_p$ is a constant of proportionality.
From the slope in Fig.\ \ref{ker} we obtain $\kappa\,=\,0.534(3)$. On
the other hand, the exponent of the thermal pseudocritical line is
$1/\beta\delta$ (see Eq.\ (\ref{pseudo})). For the 2D Ising model,
$\beta=1/8$, $\delta=15$, so that $1/\beta\delta=8/15=0.533$. The
agreement between this value and the exponent found for the Kert\'esz
line is excellent.

\begin{table}[h]
  \begin{center}{
      \begin{tabular}{|c|c|}%\hline
\hline
$h$ & $J/kT_p$\\ \hline\hline
0.00025&0.43933(3)\\ \hline
0.00050&0.43875(4) \\ \hline
0.00100&0.43786(4) \\ \hline
0.00200&0.43662(4) \\ \hline
      \end{tabular}
      }
\vskip0.5cm
\caption
{\label{thre} Percolation thresholds for the 2D Ising model
at different values of the field.}
\end{center}
\end{table}
\vskip-0.6cm

\begin{figure}[htb]
\begin{center}
\epsfig{file=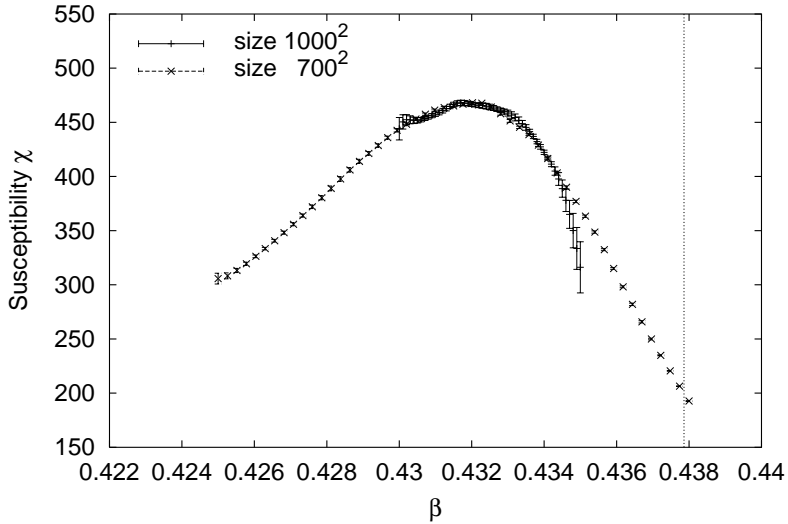,width=12cm}
\caption{\label{susc}{Susceptibility $\chi$ for the $700^2$ and
$1000^2$ lattices at $h=0.001$.}}
\end{center}
\end{figure}

We thus ask ourselves whether the Kert\'esz line and the pseudocritical
line overlap in the range of $h$ values we have chosen. 
To answer this, we determine some points of the
pseudocritical line and compare them to the corresponding points of
the Kert\'esz line. Fig.\ \ref{susc} shows the susceptibility as a
function of $\beta$ for $h=0.001$, calculated on the two largest
lattices we have studied. The curves are obtained by interpolating the
data points using the density of state method \cite{DSM}. The Figure
clearly indicates that there is no noticeable change of the peak between
$700^2$ to $1000^2$, so that we are effectively at the infinite volume
limit. The position of the peak in the Figure, however, clearly differs
from the corresponding percolation point at $h=0.001$, indicated by
the dotted line. 

To complete the comparison, we determine the positions of the
susceptibility peaks for two other values of the field, $h=0.0005$
and $h=0.002$. As above, we find clear discrepancies with the
corresponding points of the Kert\'esz line; the results are included
in Fig.\ \ref{ker}, where it is also seen that the power law
of Eq.\ (\ref{pseudo})
\be
t_{\chi} = a_{\chi} h^{1/\beta\delta}, ~~a_{\chi}={\rm const.}
\label{a-chi}
\ee
continues to be valid with great precision. We conclude that for the 2D
Ising model the functional dependence of the reduced temperature $t$ on
the field $h$ for $h{\rightarrow}0$ is 
for the Kert\'esz line and for the pseudocritical line
described by the same function, determined by the thermal critical
exponent $1/\beta\delta$; but the two curves do not coincide, since
$a_p \not= a_{\chi}$, as is evident in Fig.\ \ref{ker}.

We now turn to the corresponding study of the 3D $O(2)$ model.
Here we have determined four points of the Kert\'esz line, for
external field $h$ values 0.001, 0.00050, 0.00025, and 0.000125, using
again the cluster update \cite{nieder} as Monte Carlo algorithm,
on lattices of size $30^3$, $40^3$, $50^3$, $60^3$, $70^3$, $80^3$,
$100^3$. The clusters we considered are those used in the Wolff
algorithm with no external field (see \cite{tere,wolff}).

\begin{figure}[htb]
\begin{center}
\epsfig{file=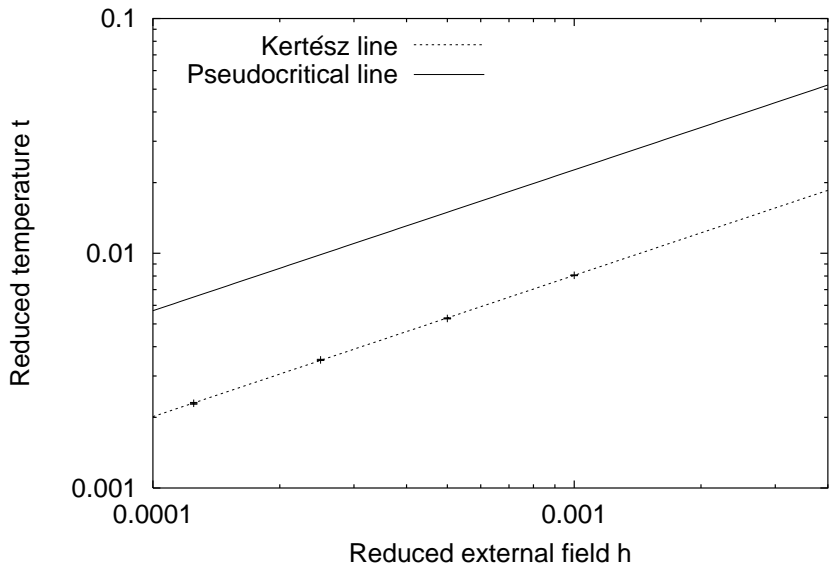,width=13cm}
\caption{\label{ker2}{Kert\'esz line and pseudocritical line for the 3D
$O(2)$ model.}}
\end{center}
\end{figure}

The positions of the percolation points, determined as before by the
crossing of the percolation cumulant curves, are listed in Table
\ref{o2tab}. As for the Ising model, we have checked that the universality
class of the exponents of each percolation transition is that of
random percolation, here in three dimensions. We show these points
in Fig. \ref{ker2}, again plotting the $h$-dependence of the reduced
percolation temperature $t_p\,=\,(T_p-T_c)/T_c$. The critical
temperature $T_c$ of the magnetization transition for the 3D $O(2)$
model without field was found to be $J/kT_c=0.454165$ in \cite{has}.

\begin{table}[h]
  \begin{center}{
      \begin{tabular}{|c|c|}%\hline
\hline
$h$ & $J/kT_p$\\ \hline\hline
0.000125&0.45312(2)\\ \hline
0.000250&0.45257(2)\\ \hline
0.000500&0.45178(3)\\ \hline
0.001000&0.45053(3)\\ \hline
      \end{tabular}
      }
\vskip0.6cm
\caption
{\label{o2tab} Percolation thresholds for the 3D $O(2)$ model
at different values of the field.}
\end{center}
\end{table}
\vskip-0.5cm

Again, the straight line in the log-log plot clearly indicates the
power law behaviour of the Kert\'esz line. The slope of the straight
line is $0.601(3)$, in good agreement with the exponent of the
pseudocritical line $1/\beta\delta=0.5995(6)$, obtained by using
the $O(2)$ thermal exponents values determined in \cite{hasen}.

To compare the two lines directly, we make use of the results of a
recent numerical determination of the $O(2)$ pseudocritical line \cite{enge}.
The results are included in Fig.\ \ref{ker2}. The two lines are parallel
in the log-log plot, since the exponents of $h$ agree, but they do not
coincide.

We have shown that the Kert\'esz lines of the 2D Ising model and
the 3D $O(2)$ model are in the limit of small fields described by
the same functions as the corresponding thermal pseudocritical lines,
i.e., a power law with exponent $1/\beta\delta$. This feature, which is
likely to be general, suggests a relationship between the geometrical
percolation singularities which still occur for $H{\neq}0$ and the
remaining non-singular thermal properties of the system. 
In both models
we have found that the Kert\'esz line does not coincide with the
pseudocritical line even in the scaling region we have explored. 
The difference in the behaviour of the two lines
could well be a consequence of using
the conventional $h=0$ bond weight (\ref{CK}) also for $h \not= 0$;
introducing a dependence on the field $h$ into the Coniglio-Klein
factor might well make the two lines coincide. Simple expressions of the
form $1-\exp[-2\beta(1+h)]$ would lead to a Kert\'esz line which goes
to infinity for $h\,\rightarrow\infty$, so that an eventual
interpretation of the pseudocritical line as a curve of percolation
points is not yet excluded. Further work on this is in progress.

\bigskip

\noindent{\bf \large Acknowledgements}

\bigskip

It is a pleasure to thank J. Engels for helpful discussions.
We would also like to thank the TMR network ERBFMRX-CT-970122 and
the DFG Forschergruppe Ka 1198/4-1 for financial support.

\end{document}